\documentclass[12pt]{article}
\usepackage{amssymb}
\usepackage{amsmath}
\usepackage{amsfonts}
\usepackage{epsfig}
\usepackage{graphicx}

\begin{document}

\title{Signal recognition and adapted filtering by non-commutative tomography%
}
\author{\textbf{Carlos Aguirre} \thanks{%
e-mail: carlos.aguirre@uam.es} \\
{\small GNB, Escuela Polit\'{e}cnica Superior, }\\
{\small \ Universidad Autonoma de Madrid, Campus de Cantoblanco, }\\
{\small \ Ctra de Colmenar Km 16, 28049 Madrid, Spain} \and \textbf{R.
Vilela Mendes} \thanks{%
e-mail: rvilela.mendes@gmail.com} \\
{\small CMAF, Complexo Interdisciplinar,}\\
{\small \ Av. Gama Pinto 2, 1649-003 Lisboa, Portugal}\\
{\small IPFN, Instituto Superior T\'{e}cnico, }\\
{\small Av. Rovisco Pais 1, 1049-001 Lisboa}}
\date{ }
\maketitle

\begin{abstract}
Tomograms, a generalization of the Radon transform to arbitrary pairs of
non-commuting operators, are positive bilinear transforms with a rigorous
probabilistic interpretation which provide a full characterization of the
signal and are robust in the presence of noise. Tomograms based on the
time-frequency operator pair, were used in the past for component separation
and denoising. Here we show how, by the construction of an operator pair
adapted to the signal, meaningful information with good time resolution is
extracted even in very noisy situations.
\end{abstract}


\section{Introduction}

\subsection{Integral transforms: linear, bilinear and tomograms}

Integral transforms \cite{handbook} \cite{BWolf79} are very useful for
signal processing in communications, engineering, medicine, physics, etc.
Linear and bilinear transforms have been used. Among the linear transforms,
Fourier \cite{Fourier1888} and wavelets \cite{Combes90} {\cite{Daubechies90} 
\cite{Chui92} are the most popular. Among the bilinear ones, the
Wigner--Ville quasidistribution \cite{Wigner32} \cite{Ville48} is the most
commonly used to provide information in the joint time--frequency domain. A
joint time--frequency description of signals is indeed important, because in
many applications (biomedical, seismic, radar, etc.) the signals are of
finite (sometimes very short) duration. However, the oscillating cross-terms
in the Wigner--Ville and other quasidistributions \cite{Cohen1} \cite{Cohen2}
\cite{Gabor} make the interpretation of these transforms a difficult matter.
Even when the average of the cross-terms is small, their amplitude may be
greater than the signal in time--frequency regions that carry no physical
information.}

{The difficulties with the physical interpretation of quasidistributions
arise from the fact that time and frequency correspond to two noncommutative
operators. Hence a joint probability density can never be defined.} Even in
the case of positive quasiprobabilities like the Husimi--Kano function \cite%
{Husimi} \cite{Kano}, an interpretation as a joint probability distribution
is also not possible because the two arguments of the function are not
simultaneously measurable random variables.

Recently, a new type of strictly positive bilinear transform has been
proposed \cite{MendesPLA} \cite{MankoJPA} \cite{RVMJRLR}, called \textit{%
tomogram}, which is a generalization of the Radon transform \cite{Radon} to
arbitrary noncommutative pairs of operators. The Radon--Wigner transform 
\cite{Radon1} \cite{Radon2} is a particular case of such noncommutative
tomography technique. The tomograms are strictly positive probability
densities, provide a full characterization of the signal and are robust in
the presence of noise.

A unified framework to characterize linear transforms, quasidistributions
and tomograms was developed in Ref.\cite{MankoJPA}. To fix notation we
briefly review it here. Signals $f(t)$ are considered as vectors $\left\vert
f\right\rangle $ in a subspace $\mathcal{N}$ of a Hilbert space $\mathcal{H}$
with dual space $\mathcal{N}^{\ast }$. Then a family of unitary operators $%
U\left( \alpha \right) =e^{iB\left( \alpha \right) }$, $\alpha $ being a
label $\left\{ \alpha \in I,I\subset \mathbb{R}^{n}\right\} $, is defined on 
$\mathcal{N}^{\ast }$. Using a ket-bra notation we denote $\left\vert
f\right\rangle \in \mathcal{N}$ and $\left\langle f\right\vert \in \mathcal{N%
}^{\ast }$. In this setting three types of integral transforms are
constructed. Let $h\in \mathcal{N}^{\ast }$ be a reference vector and let $U$
be such that the linear span of $\left\{ U(\alpha )h\in \mathcal{N}^{\ast
}:\alpha \in I\right\} $ is dense in $\mathcal{N}^{\ast }$. In $\left\{
U(\alpha )h\right\} $, a complete set of vectors can be chosen to serve as
basis.

\textbf{1 - Linear transforms} 
\begin{equation}
W_{f}^{(h)}(\alpha )=\langle U\left( \alpha \right) h\mid f\rangle
\label{2.1}
\end{equation}

\textbf{2 - Quasi-distributions} 
\begin{equation}
Q_{f}(\alpha )=\langle U\left( \alpha \right) f\mid f\rangle  \label{2.2}
\end{equation}

\textbf{3 - Tomograms}

Given an unitary $U\left( \alpha \right) =e^{iB\left( \alpha \right) }$, $%
B\left( \alpha \right) $ has the spectral projection $B\left( \alpha \right)
=\int XP\left( X\right) dX$. Let%
\begin{equation*}
P\left( X\right) \circeq \left\vert X\right\rangle \left\langle X\right\vert
\end{equation*}%
be the projector\footnote{%
Another convenient notation for the projector on a generalized eigenvector
of $B\left( \alpha \right) $ with eigenvalue $X$ is%
\begin{equation*}
\delta \left( B\left( \alpha \right) -X\right) \circeq P\left( X\right)
\end{equation*}%
} on the (generalized) eigenvector $\left\langle X\right\vert \in \mathcal{N}%
^{\ast }$ of $B\left( \alpha \right) $. The tomogram is 
\begin{equation}
M_{f}^{(B)}(X)=\left\langle f\right\vert P\left( X\right) \left\vert
f\right\rangle =\left\langle f\right\vert \left. X\right\rangle \left\langle
X\right. \left\vert f\right\rangle =\left\vert \left\langle X\right.
\left\vert f\right\rangle \right\vert ^{2}  \label{2.3}
\end{equation}%
The tomogram $M_{f}^{(B)}(X)$ is the squared amplitude of the projection of
the signal $\left\vert f\right\rangle \in \mathcal{N}$ on the eigenvector $%
\left\langle X\right\vert \in \mathcal{N}^{\ast }$ of the operator $B\left(
\alpha \right) $. Therefore it is positive. For normalized $\mid f\rangle $, 
\begin{equation*}
\langle f\mid f\rangle =1
\end{equation*}%
the tomogram is normalized 
\begin{equation}
\int M_{f}^{(B)}\left( X\right) \,dX=1  \label{2.5}
\end{equation}%
and may be interpreted as a probability distribution on the set of
generalized eigenvalues of $B\left( \alpha \right) $, that is, as the
probability distribution for the random variable $X$ corresponding to the
observable defined by the operator $B\left( \alpha \right) $.

The tomogram is a homogeneous function 
\begin{equation}
M_{f}^{(B/p)}(X)=|p|M_{f}^{(B)}(pX)  \label{2.6}
\end{equation}

\textbf{Examples:}

If $U\left( \alpha \right) $ is unitary generated by $B_{F}\left( 
\overrightarrow{\alpha }\right) =\alpha _{1}t+i\alpha _{2}\frac{d}{dt}$ and $%
h$ is a (generalized) eigenvector of the time-translation operator the
linear transform $W_{f}^{(h)}(\alpha )$ is the Fourier transform. For the
same $B_{F}\left( \overrightarrow{\alpha }\right) $\textbf{,} the
quasi-distribution\textbf{\ }$Q_{f}(\alpha )$ is the ambiguity function and
the Wigner--Ville transform {\cite{Wigner32} \cite{Ville48}} is the
quasi-distribution $Q_{f}(\alpha )$ for the following $B-$operator 
\begin{equation}
B^{(WV)}(\alpha _{1},\alpha _{2})=-i2\alpha _{1}\frac{d}{dt}-2\alpha _{2}t+%
\frac{\pi \left( t^{2}-\frac{d^{2}}{dt^{2}}-1\right) }{2}\,  \label{2.7}
\end{equation}

The wavelet transform is $W_{f}^{(h)}(\alpha )$ for $B_{W}\left( 
\overrightarrow{\alpha }\right) =\alpha _{1}D+i\alpha _{2}\frac{d}{dt}$, $D$
being the dilation operator $D=-\frac{1}{2}\left( it\frac{d}{dt}+i\frac{d}{dt%
}t\right) $. The wavelets $h_{s,\,\tau }(t)$ are kernel functions generated
from a basic wavelet $h(\tau )$ by means of a translation and a rescaling $%
(-\infty <\tau <\infty ,$ $s>0)$: 
\begin{equation}
h_{s,\,\tau }(t)=\frac{1}{\sqrt{s}}\,h\left( \frac{t-\tau }{s}\right)
\label{2.8}
\end{equation}%
using the operator 
\begin{equation}
U^{(A)}(\tau ,s)=\exp (i\tau \hat{\omega})\exp (i\log \,sD),  \label{2.9}
\end{equation}%
\begin{equation}
h_{s,\tau }(t)=U^{(A)\dagger }(\tau ,s)h(t).  \label{2.10}
\end{equation}%
The Bertrand transform \cite{BerBerJMP} \cite{Baran} is the
quasi-distribution $Q_{f}(\alpha )$ for $B_{W}\smallskip $. Linear, bilinear
and tomogram transforms are related to one another (see \cite{MankoJPA}).

\subsection{Tomograms: Some examples}

As shown above, tomograms are obtained from projections on the eigenstates
of the $B$ operators. These operators may be linear combinations of
different (commuting or noncommuting) operators, 
\begin{equation*}
B=\mu O_{1}+\nu O_{2}
\end{equation*}%
meaning that the tomogram explores the signal along lines in the plane $%
\left( O_{1},O_{2}\right) $. For example for 
\begin{equation*}
B\left( \mu ,\nu \right) =\mu t+\nu \omega =\mu t+i\nu \frac{d}{dt}
\end{equation*}%
the tomogram is the expectation value of a projection operator with support
on a line in the time--frequency plane 
\begin{equation}
X=\mu t+\nu \omega  \label{3.2}
\end{equation}%
Therefore, $M_{f}^{(S)}\left( X,\mu ,\nu \right) $ is the marginal
distribution of the variable $X$ along this line in the time--frequency
plane. The line is rotated and rescaled when one changes the parameters $\mu 
$ and $\nu $. In this way, the whole time--frequency plane is sampled and
the tomographic transform contains all the information on the signal.
Instead of marginals collected along straight lines on the time--frequency
plane, one may use other curves to sample this space \cite{MankoJPA}.

Tomograms associated to the generators of the conformal group:

\textit{Time-frequency} 
\begin{equation}
B_{1}=\mu t+i\nu \frac{d}{dt}  \label{3.5}
\end{equation}

\textit{Time-scale} 
\begin{equation}
B_{2}=\mu t+i\nu \left( t\frac{d}{dt}+\frac{1}{2}\right)  \label{3.6}
\end{equation}

\textit{Frequency-scale} 
\begin{equation}
B_{3}=i\mu \frac{d}{dt}+i\nu \left( t\frac{d}{dt}+\frac{1}{2}\right)
\label{3.7}
\end{equation}

\textit{Time-conformal} 
\begin{equation}
B_{4}=\mu t+i\nu \left( t^{2}\frac{d}{dt}+t\right)  \label{3.8}
\end{equation}

The construction of the tomograms reduces to the calculation of the
generalized eigenvectors of each one of the $B_{i}$ operators

$B_{1}\psi _{1}\left( \mu ,\nu ,t,X\right) =X\psi _{1}\left( \mu ,\nu
,t,X\right) $%
\begin{equation}
\psi _{1}\left( \mu ,\nu ,t,X\right) =\exp i\left( \frac{\mu t^{2}}{2\nu }-%
\frac{tX}{\nu }\right)  \label{3.9}
\end{equation}
with normalization 
\begin{equation}
\int dt\psi _{1}^{*}\left( \mu ,\nu ,t,X\right) \psi _{1}\left( \mu ,\nu
,t,X^{\prime }\right) =2\pi \nu \delta \left( X-X^{\prime }\right)
\label{3.10}
\end{equation}

$B_{2}\psi _{2}\left( \mu ,\nu ,t,X\right) =X\psi _{2}\left( \mu ,\nu
,t,X\right) $%
\begin{equation}
\psi _{2}\left( \mu ,\nu ,t,X\right) =\frac{1}{\sqrt{\left| t\right| }}\exp
i\left( \frac{\mu t}{\nu }-\frac{X}{\nu }\log \left| t\right| \right)
\label{3.11}
\end{equation}
\begin{equation}
\int dt\psi _{2}^{*}\left( \mu ,\nu ,t,X\right) \psi _{2}\left( \mu ,\nu
,t,X^{\prime }\right) =4\pi \nu \delta \left( X-X^{\prime }\right)
\label{3.12}
\end{equation}

$B_{3}\psi _{3}\left( \mu ,\nu ,\omega ,X\right) =X\psi _{3}\left( \mu ,\nu
,\omega ,X\right) $%
\begin{equation}
\psi _{3}\left( \mu ,\nu ,t,X\right) =\exp \left( -i\right) \left( \frac{\mu 
}{\nu }\omega -\frac{X}{\nu }\log |\omega |\right)  \label{3.13}
\end{equation}
\begin{equation}
\int d\omega \psi _{1}^{*}\left( \mu ,\nu ,\omega ,X\right) \psi _{1}\left(
\mu ,\nu ,\omega ,X^{\prime }\right) =2\pi \nu \delta \left( X-X^{\prime
}\right)  \label{3.14}
\end{equation}

$B_{4}\psi _{4}\left( \mu ,\nu ,t,X\right) =X\psi _{4}\left( \mu ,\nu
,t,X\right) $%
\begin{equation}
\psi _{4}\left( \mu ,\nu ,t,X\right) =\frac{1}{\left| t\right| }\exp i\left( 
\frac{X}{\nu t}+\frac{\mu }{\nu }\log \left| t\right| \right)  \label{3.15}
\end{equation}
\begin{equation}
\int dt\psi _{4}^{*}\left( \mu ,\nu ,t,s\right) \psi _{4}\left( \mu ,\nu
,t,s^{\prime }\right) =2\pi \nu \delta \left( s-s^{\prime }\right)
\label{3.16}
\end{equation}

Then the tomograms are:

\textit{Time-frequency tomogram} 
\begin{equation}
M_{1}\left( \mu ,\nu ,X\right) =\frac{1}{2\,\pi |\nu |}\left| \int \exp %
\left[ \frac{i\mu t^{2}}{2\,\nu }-\frac{itX}{\nu }\right] f(t)\,dt\right|
^{2}  \label{3.17}
\end{equation}

\textit{Time-scale tomogram} 
\begin{equation}
M_{2}(\mu ,\nu ,X)=\frac{1}{2\pi |\nu |}\left| \int dt\,\frac{f(t)}{\sqrt{|t|%
}}e^{\left[ i\left( \frac{\mu }{\nu }t-\frac{X}{\nu }\log |t|\right) \right]
}\right| ^{2}  \label{3.18}
\end{equation}

\textit{Frequency-scale tomogram} 
\begin{equation}
M_{3}(\mu ,\nu ,X)=\frac{1}{2\pi |\nu |}\left| \int d\omega \,\frac{f(\omega
)}{\sqrt{|\omega |}}e^{\left[ -i\left( \frac{\mu }{\nu }\omega -\frac{X}{\nu 
}\log |\omega |\right) \right] }\right| ^{2}  \label{3.19}
\end{equation}
$f(\omega )$ being the Fourier transform of $f(t)$

\textit{Time-conformal tomogram} 
\begin{equation}
M_{4}(\mu ,\nu ,X)=\frac{1}{2\pi |\nu |}\left\vert \int dt\,\frac{f(t)}{|t|}%
e^{\left[ i\left( \frac{X}{\nu t}+\frac{\mu }{\nu }\log |t|\right) \right]
}\right\vert ^{2}  \label{3.20}
\end{equation}%
The tomograms $M_{1},M_{2}$ and $M_{4}$ interpolate between the (squared)
time signal ($\nu =0$) and its projection on the $\psi _{i}\left( \mu ,\nu
,t,X\right) $ functions for $\mu =0$.

In a similar way, tomograms may be constructed for any operator of the
general type 
\begin{equation*}
B_{4}=\mu t+i\nu \left( g\left( t\right) \frac{d}{dt}+\frac{1}{2}\frac{%
dg\left( t\right) }{dt}\right)
\end{equation*}%
the generalized eigenvectors being 
\begin{equation*}
\psi _{g}\left( \mu ,\nu ,t,X\right) =\left\vert g\left( t\right)
\right\vert ^{-1/2}\exp i\left( -\frac{X}{\nu }\int^{t}\frac{ds}{g\left(
s\right) }+\frac{\mu }{\nu }\int^{t}\frac{sds}{g\left( s\right) }\right)
\end{equation*}

When dealing with finite-time signals and finite-time tomograms some
normalization modifications are needed. For example, for a time-frequency
tomogram,\ instead of (\ref{3.17}), we consider the finite-time tomogram,
for a signal defined from $t_{0}$ to $t_{0}+T$%
\begin{equation}
M_{1}(\theta ,X)=\left\vert \int_{t_{0}}^{t_{0}+T}\ f^{\ast }(t)\psi
_{\theta ,X}^{(1)}\left( t\right) \,dt\right\vert ^{2}=\left\vert <f,\psi
^{(1)}>\right\vert ^{2}  \label{4.1}
\end{equation}%
with 
\begin{equation}
\psi _{\theta ,X}^{(1)}\left( t\right) =\frac{1}{\sqrt{T}}\exp \left( \frac{%
i\cos \theta }{2\sin \theta }\,t^{2}-\frac{iX}{\sin \theta }\,t\right)
\label{4.2}
\end{equation}%
and $\mu =\cos \theta ,\nu =\sin \theta $. $\theta $ is a parameter that
interpolates between the time and the frequency operators, running from $0$
to $\pi /2$ whereas $X$ is allowed to be any real number. An orthonormalized
set of $\psi _{\theta ,X}^{(1)}\left( t\right) $ vectors is obtained by
choosing the sequence%
\begin{equation}
X_{n}=X_{0}+\frac{2n\pi }{T}\sin \theta \hspace{2cm}n\in \mathbb{Z}
\label{4.2a}
\end{equation}

Likewise for the finite-time time-scale tomogram $M_{2}(\mu ,\nu ,X)$ (Eq.%
\ref{3.18}) and the finite-time time-conformal tomogram $M_{4}(\mu ,\nu ,X)$
(Eq.\ref{3.20}): 
\begin{equation}
M_{2}\left( \theta ,X\right) =\left\vert \int_{t_{0}}^{t_{0}+T}\ f^{\ast
}(t)\psi _{\theta ,X}^{(2)}\left( t\right) \,dt\right\vert ^{2}=\left\vert
<f,\psi ^{(2)}>\right\vert ^{2}  \label{4.10}
\end{equation}%
\begin{equation}
\psi _{\theta ,X}^{(2)}\left( t\right) =\frac{1}{\sqrt{\log \left\vert
t_{0}+T\right\vert -\log \left\vert t_{0}\right\vert }}\frac{1}{\sqrt{%
\left\vert t\right\vert }}\exp i\left( \frac{\cos \theta }{\sin \theta }\,t-%
\frac{X}{\sin \theta }\,\log \left\vert t\right\vert \right)  \label{4.11}
\end{equation}%
\begin{equation}
X_{n}=X_{0}+\frac{2n\pi }{\log \left\vert t_{0}+T\right\vert -\log
\left\vert t_{0}\right\vert }\sin \theta \hspace{2cm}n\in \mathbb{Z}
\label{4.12}
\end{equation}%
and 
\begin{equation}
M_{4}(\theta ,X)=\left\vert \int_{t_{0}}^{t_{0}+T}\ f^{\ast }(t)\psi
_{\theta ,X}^{(4)}\left( t\right) \,dt\right\vert ^{2}=\left\vert <f,\psi
^{(4)}>\right\vert ^{2}  \label{4.13}
\end{equation}%
\begin{equation}
\psi _{\theta ,X}^{(4)}\left( t\right) =\sqrt{\frac{t_{0}\left(
t_{0}+T\right) }{T}}\frac{1}{\left\vert t\right\vert }\exp i\left( \frac{%
\cos \theta }{\sin \theta }\,\,\log \left\vert t\right\vert +\frac{X}{t\sin
\theta }\right)  \label{4.14}
\end{equation}%
\begin{equation}
X_{n}=X_{0}+\frac{t_{0}\left( t_{0}+T\right) }{T}2\pi n\sin \theta \hspace{%
2cm}n\in \mathbb{Z}  \label{4.15}
\end{equation}

\section{Applications of tomograms: Denoising, and component separation}

Most natural and man-made signals are nonstationary and have a
multicomponent structure. Therefore separation of its components is an issue
of great technological relevance. However, the concept of signal component
is not uniquely defined. The notion of \textit{component} depends as much on
the observer as on the observed object. When we speak about a component of a
signal we are in fact referring to a particular feature of the signal that
we want to emphasize. For signals that have distinct features both in time
and in the frequency domain, the time-frequency tomogram is an appropriate
tool.

Consider finite-time tomograms as in (\ref{4.1}). For all different $\theta $%
's the $U({\theta })$, of which $B\left( \theta \right) $ is the
self-adjoint generator, are unitarily equivalent operators, hence all the
tomograms share the same information.

First we would select a subset $X_{n}$ in such a way that the corresponding
family $\left\{ \psi _{\theta ,X_{n}}^{(1)}\left( t\right) \right\} $ is
orthogonal and normalized, 
\begin{equation}
<\psi _{\theta ,X_{n}}^{(1)}\psi _{\theta ,X_{m}}^{(1)}>=\delta _{m,n}
\label{4.3}
\end{equation}%
This is the sequence listed in (\ref{4.2a}), where $X_{0}$ is freely chosen
(in general we take $X_{0}=0$). We then consider the projections of the
signal $f(t)$ on this set of $\psi _{\theta ,X_{n}}^{(1)}$ vectors%
\begin{equation}
c_{X_{n}}^{\theta }(f)=<f,\psi _{\theta ,X_{n}}^{(1)}>  \label{4.5}
\end{equation}

\textit{Denoising} consists in eliminating the $c_{X_{n}}^{\theta }(f)$ such
that 
\begin{equation}
\left| c_{X_{n}}^{\theta }(f)\right| ^{2}\leq \epsilon  \label{4.6}
\end{equation}
for some threshold $\epsilon $. This power selective denoising is more
robust than, for example, frequency filtering which may also eliminate
important signal information.

The \textit{component separation technique} is based on the search for an
intermediate value of ${\theta }$ where a good compromise might be found
between time localization and frequency information. This is achieved by
selecting subsets $\mathcal{F}_{k}$ of the $X_{n}$ and reconstructing
partial signals ($k$-components) by restricting the sum to 
\begin{equation}
f_{k}(t)=\sum_{n\in \mathcal{F}_{k}}c_{X_{n}}^{\theta }(f)\psi _{\theta
,X_{n}}(t)  \label{4.7}
\end{equation}%
for each $k$.

\section{Signal detection with an adapted operator pair}

Time-frequency tomograms have been used for denoising and component
separation of finite-time signals \cite{MendesPLA} \cite{reflecto1} \cite%
{reflecto2} \cite{Clairet} \cite{Aguirre}. Time-frequency tomograms are
particularly appropriate to identify the time unfolding of the frequency
features of the signals. For example, the component separation success \cite%
{reflecto1} \cite{reflecto2} in the plasma reflectometry applications is to
a large extent due to the fact that the plasma is sampled by microwave
chirps and the basis in (\ref{4.2}) is exactly a chirp basis. This suggests
that, for other types of signals, other types of tomograms should be chosen.

In particular, if in the linear combination $B\left( \mu ,\nu \right) =\mu
t+\nu O$, one chooses an operator $O$, that is specially tuned to the
features of the signal that one wants to extract, then, by looking for the
particular values of the set $\left( \mu ,\nu \right) $ where the noise
effects might cancel out, we may separate the information of very small
signals from large noise and also obtain reliable information on the
temporal structure of the signal. This would provide a signal-adapted
filtering technique. The construction of the operator suited to particular
signals may be done by the same techniques that are used in the
bi-orthogonal decomposition \cite{Dente}.

The method for the construction of the adapted operator pair is as follows:

Consider a set of $N-$dimensional time sequences $\left\{ \overrightarrow{%
x_{1}},\cdots ,\overrightarrow{x_{k}}\right\} $, typical of the signal one
wants to detect. For a communication point of view these may be considered
as the code words that, later on, one wishes to detect in a noisy signal.
Form the $k\times N$ matrix $U\in \mathcal{M}_{k\times N}$.

\begin{equation}
U=\left( 
\begin{array}{cccc}
x_{1}(1\Delta t) & x_{1}(2\Delta t) & \ldots & x_{1}(N\Delta t) \\ 
\vdots & \vdots &  & \vdots \\ 
x_{k}(1\Delta t) & x_{k}(2\Delta t) & \ldots & x_{k}(N\Delta t)%
\end{array}%
\right)  \label{5.1}
\end{equation}%
with $k<N$ typically.

Now construct the square matrices $A=U^{T}U\in \mathcal{M}_{N\times N}$ and $%
B=UU^{T}\in \mathcal{M}_{k\times k}$. The diagonalization of $A$ provides $k$
non-zero eigenvalues $(\alpha _{1},\cdots ,\alpha _{k})$ and its
corresponding orthogonal $N-$dimensional eigenvectors $(\Phi _{1},\cdots
,\Phi _{k})$, $\Phi _{j}\in \mathbb{R}^{N}$. Correspondingly, the
diagonalization of $B$ would provide the same $k$ eigenvalues and
eigenvectors $(\Psi _{1},\cdots ,\Psi _{k})$ with $\Psi _{j}\in \mathbb{R}%
^{k}$. If needed one may obtain, by the Gram-Schmidt method, the remaining $%
N-k$ eigenvectors to span $\mathbb{R}^{N}$, which in this context are
associated to the eigenvalue zero.

The linear operator $S$ constructed from the set of typical signals is%
\begin{equation}
S=\sum_{i=1}^{k}\alpha _{i}\Phi _{i}\Phi _{i}^{t}  \label{5.2}
\end{equation}%
where $S\in \mathcal{M}_{N\times N}$.

For the tomogram we consider an operator $B\left( \mu ,\nu \right) $ of the
form%
\begin{equation}
B\left( \mu ,\nu \right) =\mu t+\nu S=\mu \left( 
\begin{array}{ccccc}
1\Delta t &  &  &  &  \\ 
& 2\Delta t &  &  &  \\ 
&  & 3\Delta t &  &  \\ 
&  &  & \ddots &  \\ 
&  &  &  & N\Delta t%
\end{array}%
\right) +\nu \sum_{i=1}^{k}\alpha _{i}\Phi _{i}\Phi _{i}^{t}  \label{5.3}
\end{equation}%
where $B\in \mathcal{M}_{N\times N}$.

The eigenvectors of each $B\left( \mu ,\nu \right) $ are the columns of the
matrix that diagonalizes it. From the projections of a signal on these
eigenvectors one constructs a tomogram adapted to the operator pair $\left(
t,S\right) $.

\subsection{Examples}

Here we present some examples of the application of the technique described
in the previous Section.

In the first example a set of $40$ random signals with pulses of duration $%
\Delta t=10$ and intensities $+1$ or $-1$ are generated. The total length of
the signal is $200$ time units. These random signals form the typical data
to which we will adapt the tomogram, by constructing the operator $S$ in (%
\ref{5.2}). Fig.1 shows $10$ signals of this type. They all vary between $+1$
and $-1$, being shifted in the figure for clarity purposes.


\begin{figure}[ht!]
\centerline{\psfig{figure=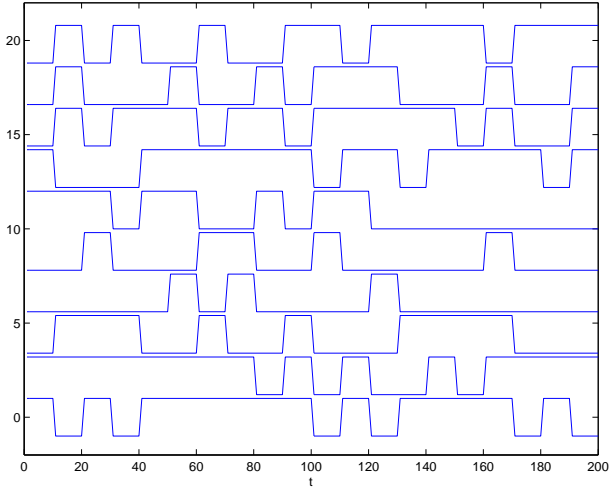,height=6.56cm}}
\caption{A set of typical signals.}
\label{fig:fig_1}
\end{figure}

Once the operator $S$ is constructed, one considers the operator%
\begin{equation*}
B\left( \theta \right) =t\cos \theta +S\sin \theta
\end{equation*}%
for which one computes the eigenbasis which is used to project the signals
to be analyzed. To a pure signal of the same type as those used to construct
the operator $S$ (in the upper left panel of Fig.2), we add Gaussian noise
(upper right panel of Fig.2). This signal is then analyzed and a tomogram
constructed for $20$ different values of $\theta $ at intervals $\Delta
\theta =\pi /40$. A contour plot of the tomogram is shown in the lower left
panel of Fig.2. As we have explained before, by inspection of the power
distribution in the tomogram one may either select the components of
intensity higher than a threshold for denoising of the whole signal or
select particular components of signal. In this case what is of interest is
to select the part of the signal that corresponds to the typical signals
used to construct $S$. This is done by selecting only the strongest
components at the region where they concentrate. In the lower right panel of
Fig.2 we show the result of projecting on the eigenvectors $185$ to $200$ at 
$\theta _{19}=19\pi /40$.


\begin{figure}[ht!]
\centerline{\psfig{figure=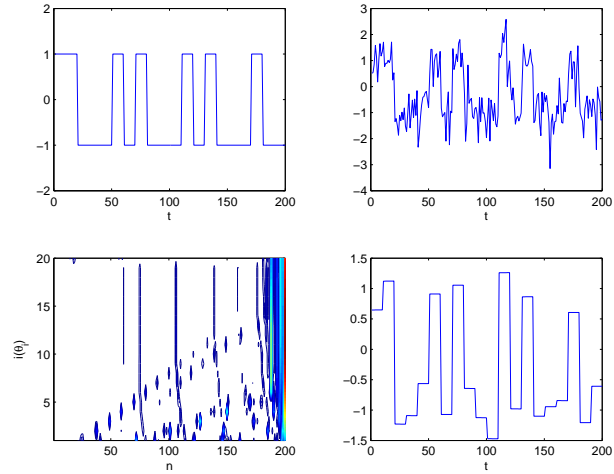,height=6.66cm}}
\caption{Signal, noisy signal, the tomogram and the projection on the
eigenvectors $185$ to $200$ at $\protect\theta =19\protect\pi /40$}
\label{fig:fig_2}
\end{figure}

One sees that the signal is reasonably reconstructed from the noisy input.
With a clipping operation at $\pm 0.5$ the reconstruction would be perfect.

For the second example we have generated, as typical signals a set of $40$
sines with random frequencies. Fig.3 displays some examples.


\begin{figure}[ht!]
\centerline{\psfig{figure=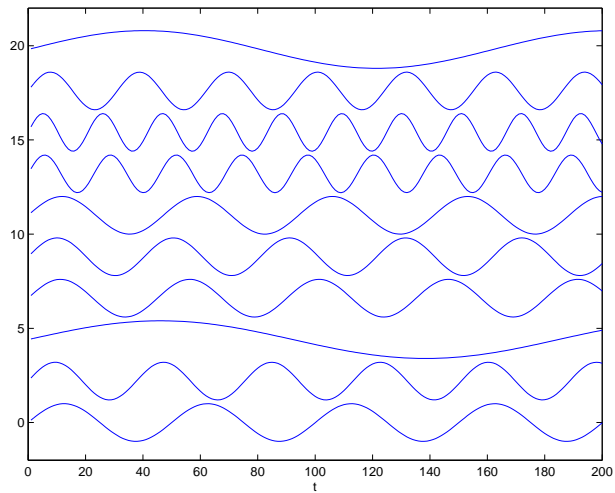,height=6.56cm}}
\caption{A set of sines with random frequencies}
\label{fig:fig_3}
\end{figure}

This case is harder because noise (or interference) is expected to contain
frequencies similar to the typical signals. Fig.4 shows the results of a
typical analysis of a signal that contains pieces of sines at different time
intervals. In the upper left panel it is the signal, in the upper right
panel the signal with Gaussian noise added, in the lower left panel the
tomogram and in the lower right panel the result of the projection on the
eigenvectors $189$ to $197$ at $\theta _{4}=15\pi /40$.


\begin{figure}[ht!]
\centerline{\psfig{figure=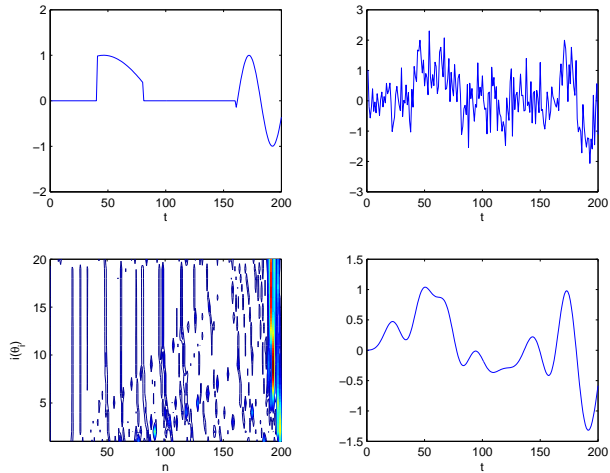,height=6.36cm}}
\caption{Signal, noisy signal, the tomogram and the projection on the
eigenvectors $189$ to $197$ at $\protect\theta =19\protect\pi /40$}
\label{fig:fig_4}
\end{figure}

The projection range that was used aimed at including all the strongest
components. Notice however that by selecting particular regions of the
tomogram one may extract particular components of the signal. For example,
Fig.5 shows the result of projection on the eigenvectors $190$ to $199$ at $%
\theta =\pi /10$.


\begin{figure}[ht!]
\centerline{\psfig{figure=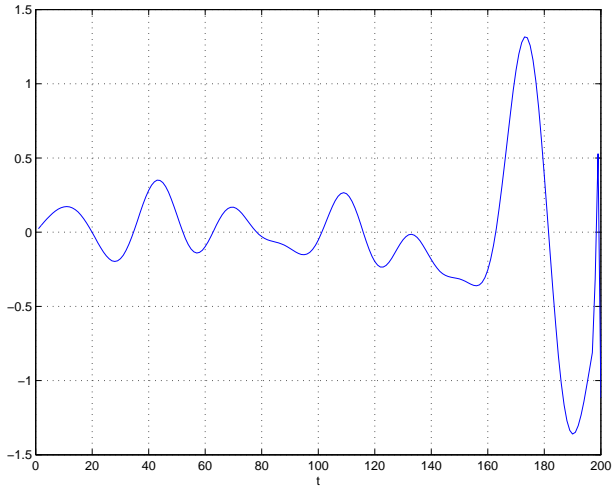,height=6.56cm}}
\caption{The projection on the eigenvectors $190$ to $199$ at $\protect%
\theta =\protect\pi /10$ which separates the second time component of the
signal}
\label{fig:fig_5}
\end{figure}

For the third example we use data obtained from the Phoenix Mars Lander \cite%
{Smith}. A dust devil is a hot whirlwind generated by a huge contrast
between the martian atmospheric air and the planet surface. Dust devils
appear in both temperature and pressure data as sudden drops with a duration
between two and three minutes. The upper left panel in Figure 6 shows some
data from the Phoenix Mars Lander covering a $2000$ seconds interval with a
sampling rate of $.5Hz$. A dust devil is clearly visible at $t\simeq 800s$
as a drop in the pressure value.

There has been several efforts to develop systematic methods to detect the
effect of dust devils on the martian atmosphere data. They are based either
on checking several ad-hoc conditions in the data \cite{Smith} or on
Field-Programmable Gate Arrays (FPGAs) \cite{Lucas}.

To use our adapted tomographic filtering method for the detection of dust
devils we have generated a set of $278$ signals that resemble the shape that
a dust devil produces on the data, that is a sudden drop of about $3\%$ from
the baseline, with different durations ranging from $60$ to $80$ time units.
The upper right panel displays several of these typical signals. Some of the
signals have been shifted up or down for representation purposes.

As in the previous examples a tomogram is constructed for $20$ different
values of $\theta $ at intervals $\Delta \theta =\pi /40$. A contour plot of
the first $999$ coefficients of the tomogram is shown in the lower left
panel of Figure 6. Coefficient $n=1000$ corresponds to the biggest
eigenvalue (and its corresponding eigenvector). This eigenvector contains
most of the energy of the signal, and is several orders of magnitude bigger
that any other coefficient, so clarity this coefficient has not been plotted
in the tomogram. By direct inspection, we observe that, besides the
coefficient $n=1000$, the strongest components concentrate close to $n=400$.
The lower right panel in figure 6 shows the projection on the eigenvectors $%
340$ to $450$ and $1000$ at $\theta =19\pi /40$. One sees that the pressure
drop produced by the dust devil is very well reconstructed and separated
from any other components present in the signal such as noise or smaller
presure variations.

\begin{figure}[ht!]
\centerline{\psfig{figure=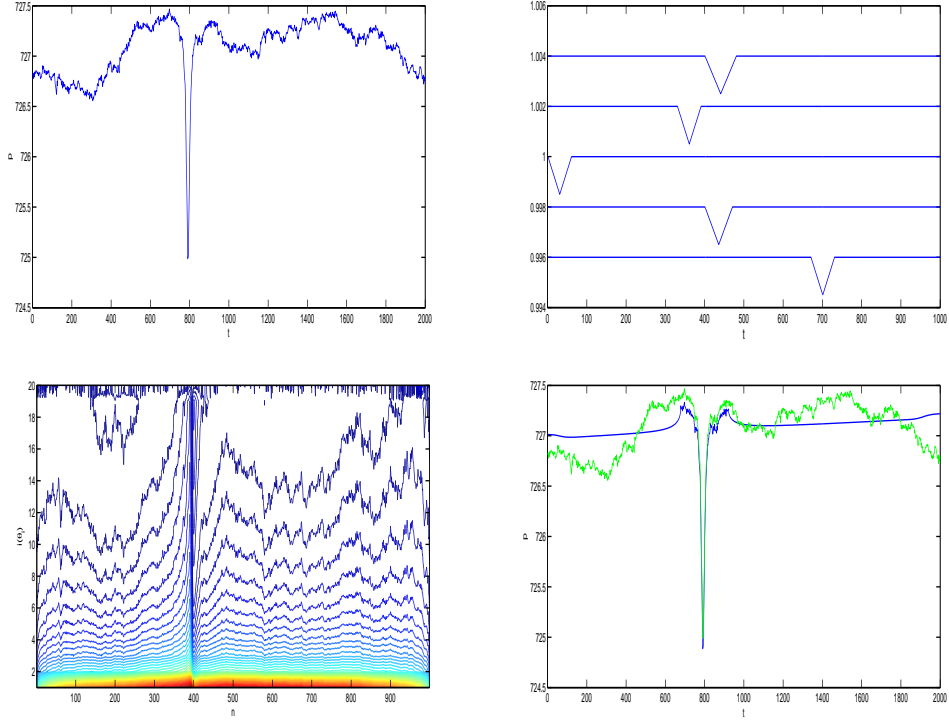,height=10cm}}
\caption{Signal, typical signals, the tomogram (coefs 1-999) and the
projection on the eigenvectors $340$ to $450$ and $1000$ at $\protect\theta %
=19\protect\pi /40$}
\label{fig:signal}
\end{figure}

As an alternative, that avoids the large value of the biggest eigenvalue, we
may shift both the typical signals and the real data to zero mean signals.
In this case there is no eigenvalue much larger than all others. The left
panel of figure 7 displays a 3D plot of the tomogram for the $1000$ coefficients
obtained with zero mean signals. We have also applied a denoising procedure,
removing the small coefficients. The right panel in figure 7 shows the
projection on the eigenvectors $340$ to $450$ at $\theta =19\pi /40$. One
sees that the pressure drop produced by the dust devil is completely
separated from any other components in the signal.

\begin{figure}[ht!]
\centerline{\psfig{figure=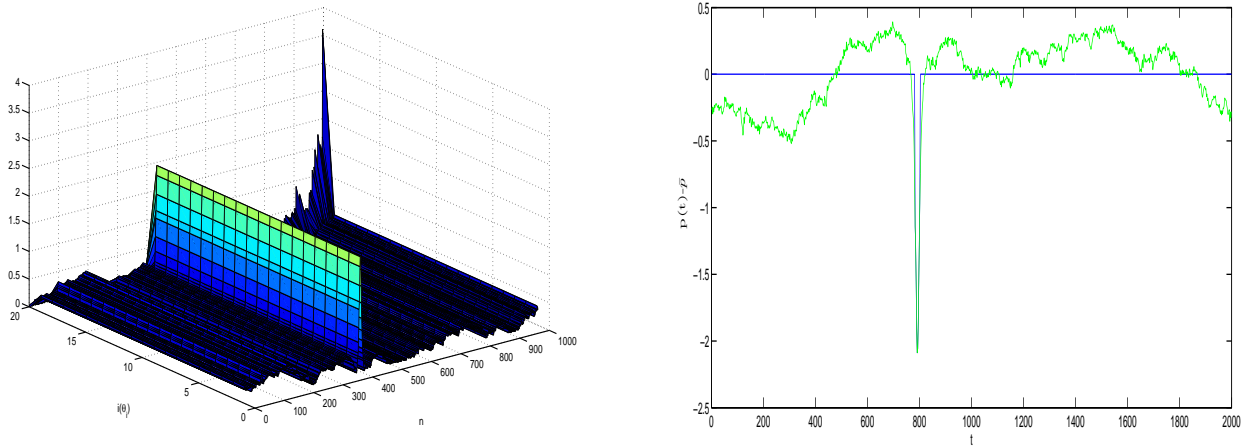,height=6.56cm}}
\caption{Tomogram for 0 mean typical signals and the projection on the
eigenvectors $340$ to $450$ at $\protect\theta =19\protect\pi /40$}
\label{fig:0mean}
\end{figure}

\section{Conclusions}

Tomograms provide a two-variable characterization of signals which, due to
its rigorous probabilistic interpretation, is robust and free of artifacts
and ambiguities. For each particular signal that one wants to analyze, the
choice of the appropriate tomogram depends not only on the signal but also
on the features that we might want to identity or emphasize.

We have developed a new family of data-driven tomograms that are combination
of time and an operator obtained from a set of typical signals specially
tuned to represent the features that one wants to extract. These adapted
tomograms provide, for noisy signals, filtering and separation of components
and features that might not be well represented by the combination of
standard operators.\\

{\bf Acknowledgments}\\

This work is partially supported by Spanish MICINN BFU2009-08473 and MEIGA
METNET PRECURSOR (AYA2011-29967-C05-02) funded by the Spanish Ministerio de
Economia y Competitividad. We would also like to thank Germ\'an Mart\'{\i}%
nez for providing the Phoenix Mars Lander data.

\end{document}